# Line geometry and electromagnetism I: basic structures

D. H. Delphenich
Kettering, OH 45440

**Abstract.** Some key notions of line geometry are recalled, along with their application to mechanics. It is then shown that most of the basic structures that one introduces in the pre-metric formulation of electromagnetism can be interpreted directly in terms of corresponding concepts in line geometry. The results are summarized in a table.

**Table of contents**

**1. Introduction.** Prior to the success of Einstein's theory of gravitation – i.e., general relativity – the pre-eminent branch of geometry was projective geometry. Indeed, the place of projective geometry in physics was much more widely known by physicists in the pre-Einstein era, mostly by way of its role in mechanics and geometrical optics. Furthermore, one's education as a geometer would have been regarded as fundamentally incomplete until one was well-versed in the concepts and methods of projective analytical geometry.

Perhaps the crowning achievement of projective geometry was Felix Klein's Erlanger Programm [**1**], which was the Habilitationsschrift that he presented in order to join the mathematics faculty at the university at Erlangen. In that thesis, he proposed that, firstly, a "geometry" should be defined by a group of transformations of a space that preserve some fundamental property or construction regarding that space (e.g., incidence, parallelism, angles, distances), and secondly, that the geometry at the top of that pyramid of subgroup inclusions would then be projective geometry. As a result, one did not get to the introduction of a metric on an affine space except by restricting projective transformations, firstly, to affine transformations, and secondly, to isometries.

However, once Einstein validated the power of Riemannian (or really, *pseudo*-Riemannian) geometry in physics, interest in projective geometry seems to have waned in both the physics and mathematics communities. Physicists were no longer expected to learn any projective geometry, while geometers only learned about some of its special topics if they intended to go into algebraic geometry, which eventually focused on the algebraic techniques more than the geometry itself. Mostly, geometers seem to have felt that the success of Riemannian geometry in general relativity was a self-sufficient reason for concentrating on it to the exclusion of other forms of geometry. To this day, the only aspect of Klein's Erlanger Programm that gets mentioned is the first part about geometries being related to groups of transformations; the fact that the top group in that "food chain" would be the projective group gets mentioned much less frequently.

As a result, one of the potential applications of projective geometry to physics that got passed over was its application to electromagnetism. That is probably because the point at which projective geometry becomes relevant to electromagnetism is when one considers the formulation of that field theory in terms of exterior differential forms, which came later in the Twentieth Century than general relativity, and is still a source of reluctance and skepticism in the physics community to this day.

The key to relating projective geometry to electromagnetism is then by way of the Plücker-Klein embedding. One first observes that under the projection $\mathbb{R}^4 - 0 \to \mathbb{R}P^3$, $\mathbf{v} \mapsto [\mathbf{v}]$, a plane through the origin in $\mathbb{R}^4$ will become a line in $\mathbb{R}P^3$. One then represents a plane through origin in $\mathbb{R}^4$ by an equivalence class of decomposable bivectors $\mathbf{a} \wedge \mathbf{b}$ that differ from each other only by a non-zero scalar multiple. Thus, if one thinks of the vector space $\Lambda_2^4$ of bivectors over $\mathbb{R}^4$ as a vector space that defines a projective space $P\Lambda_2^4$ by its lines through the origin then any plane through the origin in $\mathbb{R}^4$ will be associated with a unique point in $P\Lambda_2^4$. Since the association of planes through origin in $\mathbb{R}^4$ and lines in $\mathbb{R}P^3$ is one-to-one, one sees that lines in $\mathbb{R}P^3$ are associated with unique points in $P\Lambda_2^4$. The image of that embedding of the manifold $\mathcal{L}(3)$ of lines in $\mathbb{R}P^3$ in $P\Lambda_2^4$ is a quadric hypersurface that one calls the "Klein quadric."

A similar argument also relates lines in $\mathbb{R}P^3$ to decomposable 2-forms over $\mathbb{R}^4$, since such a 2-form will annihilate a plane through the origin in $\mathbb{R}^4$.

The fact that for dimension four, a line can be described by either a decomposable bivector or a decomposable 2-form amounts to the fact that a line in $\mathbb{R}P^3$ can be expressed as either the join of two distinct points or the meet (i.e., intersection) of two distinct planes. Similarly, Plücker [**2**] regarded a line as either a one-parameter family of points or a one-parameter family of planes that all intersected a common axis; i.e., a pencil of planes.

Since the basic fields of electromagnetism can be expressed as bivector fields and 2-forms, which are usually decomposable, moreover, the association of electromagnetism with line geometry becomes immediate. Furthermore, any 2-form, such as the electromagnetic field strength 2-form will define a "linear complex," while linear, non-dispersive electromagnetic constitutive laws will define "correlations," and thus, "quadratic line complexes." The existence of wave-like solutions to Maxwell's equations (when expressed in "pre-metric" form) brings about a reduction of the group of projective transformations of $\mathbb{R}P^3$ (which will take lines to lines) to a subgroup that is isomorphic to the special Lorentz group. However, the group-theoretic aspects of line geometry are somewhat involved in their own right and will thus be deferred to the second part of this two-part article.

Hence, in this first part, we shall confine ourselves to introducing the basic concepts of projective geometry, line geometry, and pre-metric electromagnetism that relate to establishing the basic association of electromagnetic structures with line-geometric ones. Since the basic fields of electromagnetism also relate to distributions of forces and moments in spacetime, we also point out some of the earlier associations between line geometry and mechanics, as well as the fact that the decomposition of spacetime into time and space can be related to projective geometry in a way that relates directly to the decomposition of bivector fields and 2-forms into electric and magnetic parts. The ultimate result of this analysis is then summarized in a table of associations between the two disciplines.

## 2. Projective geometry.

We shall briefly summarize the essential concepts of projective geometry that will be applied to the geometry of lines in $\mathbb{R}P^3$. Some relevant treatments of projective analytic geometry can be found in [**3-6**], and a modern treatment that is explicitly based in the Erlanger Programm is Onishchik and Sulanke [**7**].

*a. Projective spaces.* Although projective spaces often get defined by simply the projection $V - 0 \to PV$, $\mathbf{v} \mapsto [\mathbf{v}]$ that takes any non-zero vector $\mathbf{v}$ in a vector space $V$ to the line through the origin $[\mathbf{v}]$ that it generates, actually, this definition is somewhat misleading. That is because the scope of projective geometry is actually *larger* than that of affine geometry, not *smaller*, as one would suspect from the fact that there is a reduction of the dimension as a result of the projection. Indeed, the best way of looking at the relationship between affine and projective geometry is to think of a projective space as something that is obtained by *adding* a (projective) hyperplane at infinity to an affine space. That is, one "completes" the affine space by adding the asymptotic behavior of things "at infinity." Hence, one should think of the previous way of obtaining a projective space as something that is more related to the introduction of homogeneous coordinates than it is to the nature of projective geometry itself.

In particular, when $V$ is $\mathbb{R}^{n+1}$, one can think of the elements $(x^0, \dots, x^n)$ of that space as the *homogeneous coordinates* for the points of $\mathbb{R}\mathrm{P}^n$. However, the projection from homogeneous to *inhomogeneous* coordinates is not by way of an elementary dropping of one coordinate (which would be an "orthogonal" projection), but of dividing through by one non-zero coordinate. If that non-zero coordinate is $x^0$ then the inhomogeneous coordinates for a subset of $\mathbb{R}\mathrm{P}^n$ will be defined by:

$$X^i = \frac{x^i}{x^0}. \tag{2.1}$$

One sees that this definition will break down for all points in the hyperplane $x^0 = 0$ in $\mathbb{R}^{n+1}$. Since the $X^i$ approach infinity as $x^0$ approaches 0, and the only point of $\mathbb{R}^{n+1}$ that does not project to $\mathbb{R}\mathrm{P}^n$ is 0, one sees that the lines through the origin that lie in the hyperplane $x^0 = 0$ will still project to well-defined points of $\mathbb{R}\mathrm{P}^n$ that will represent a projective subspace of codimension one; one calls it the *hyperplane at infinity* in $\mathbb{R}\mathrm{P}^n$ ([1]), which we denote by $\mathbb{R}\mathrm{P}^{n-1}_\infty$. The remaining points in $\mathbb{R}\mathrm{P}^n$ that do not lie at infinity are called *finite* points.

The points in $\mathbb{R}^{n+1}$ whose homogeneous coordinates are of the form $(1, x^i)$ have the special property that the inhomogeneous coordinates that are associated with $x^i$ are the same as the homogeneous coordinates. Thus, in some sense the affine hyperplane $x^0 = 1$ makes a good model for the finite points of $\mathbb{R}\mathrm{P}^n$, while $x^0 = 0$ represents the points at infinity.

One sees that since we could just as well have chosen any other non-zero homogeneous coordinate to divide by:

a. It will take $n + 1$ inhomogeneous coordinate charts to cover $\mathbb{R}\mathrm{P}^n$.

b. The distinction between finite and infinite points is as arbitrary as the choice of hyperplane through the origin in $\mathbb{R}^{n+1}$.

This suggests that the direct sum decomposition $\mathbb{R}^{n+1} = \mathbb{R} \oplus \mathbb{R}^n$ is closely related to the subset partitioning of $\mathbb{R}\mathrm{P}^n$ into a set of finite points and a set of points at infinity. If the $\mathbb{R}$ summand is generated by a non-zero vector $\mathbf{t}$ then any vector $\mathbf{v} \in \mathbb{R}^{n+1}$ can be uniquely expressed in the form:

---

([1]) A possible source of confusion in projective geometry is the fact that typically one drops the adjective "projective" when referring to subspaces of projective spaces. Thus, a line, plane, etc., in a projective space is always a *projective* line, plane, etc.

$$\mathbf{v} = v^0\,\mathbf{t} + \mathbf{v}^s, \tag{2.2}$$

where $\mathbf{v}^s \in \mathbb{R}^n$.

If one chooses any $n$-frame $\{\mathbf{e}_i, i = 1, \ldots, n\}$ for $\mathbb{R}^n$ then $\mathbf{v}$ can be associated with components $(v^0, v^1, \ldots, v^n)$, and if they are treated as the homogeneous coordinates for a point in $\mathbb{R}\mathrm{P}^n$ then that suggests that the vectors $\mathbf{v}^s \in \mathbb{R}^n$ (i.e., $v^0 = 0$) will project to the points at infinity. The (finite) point $[\mathbf{t}] \in \mathbb{R}\mathrm{P}^n$ that corresponds to the vector $\mathbf{t}$ will play a special role by the fact that it is associated with the origin in $\mathbb{R}^n$, and we then denote $[\mathbf{t}]$ by $\mathcal{O}$.

In the interests of modernity, one can also characterize a projective space as a set with a "projective structure." We shall understand this term to mean the lattice ([1]) of projective subspaces of a given projective space, for which, the partial ordering is set inclusion, and the two binary operations are called "join" and "meet," which we denote by $\vee$ and $\cap$, respectively. The *join* of two subspaces is the smallest subspace that contains both of them, so it is generally bigger than their union. The *meet* of two subspaces is the largest subspace that they both contain, so it is precisely their intersection. The lattice in question also has a unique greatest element – viz., the entire space, which contains every subspace – and a unique least element – viz., the empty subset, which is contained in every subspace.

For instance, the join of two distinct points is a line, the join of a line and a distinct point is plane, and the join of two lines in a space of dimension $n > 2$ might be a line (if they are coincident), a plane (if they intersect), or a three-dimensional subspace (if they are skew). By contrast, the meet of two planes in a space of dimension $n > 2$ can be empty (but only if $n > 3$), a point (only if $n > 3$), a line (if they are distinct) or a plane (if they are coincident). The meet of two lines can be empty (if they are skew), a point (if they are distinct), or a line (if they are coincident). The meet of any point with any subspace of dimension greater than zero will be empty if the point is not incident on the subspace and the point itself if it is.

The concept of *incidence* that we just invoked is basically just a symmetrization of the concept of inclusion. That is, one subspace $A$ is incident on another $B$ iff either $A$ is a subspace of $B$ or vice versa. The reason for the symmetrization relates to the concept of duality, which we will discuss next. In the eyes of Klein's Erlanger Programm, incidence is the fundamental concept that defines projective geometry, just as parallelism defines affine geometry, and distance defines metric geometry.

*b. Correlations.* Just as the vector space $\mathbb{R}^{n+1}$ has a dual space $\mathbb{R}^{n+1*}$, which is composed of all linear functionals on $\mathbb{R}^{n+1}$, by considering all lines through the origin in

([1]) A partially-ordered set that has been given two binary operations, which are first introduced as the "greatest lower bound" and "least upper bound," is called a *lattice* [**8**].

$\mathbb{R}^{n+1*}$, one will define the projective space $\mathbb{R}P^{n*}$ that is dual to $\mathbb{R}P^n$. The subspaces of $\mathbb{R}P^{n*}$ are associated with subspaces of $\mathbb{R}P^n$ by the fact that any linear functional $\alpha$ on $\mathbb{R}^{n+1}$ will define a hyperplane Ann($\alpha$) = {all $\mathbf{v} \in \mathbb{R}^{n+1}$ | $\alpha(\mathbf{v}) = 0$} in $\mathbb{R}^{n+1}$ that one calls *annihilating hyperplane* of $\alpha$. Thus, a point $[\alpha]$ in $\mathbb{R}P^{n*}$ represents a (projective) hyperplane in $\mathbb{R}P^n$, and, more generally, a $k$-dimensional subspace of $\mathbb{R}P^{n*}$ will represent an $n-k-1$-dimensional subspace of $\mathbb{R}P^n$. It is important to not that when $n = 3$, a line in $\mathbb{R}P^{3*}$ is associated with a line in $\mathbb{R}P^3$, as well.

The incidence of points and hyperplanes in $\mathbb{R}P^n$ is closely related to the canonical bilinear pairing of $\mathbb{R}^{n+1}$ with $\mathbb{R}^{n+1*}$ that takes a pair $(\alpha, \mathbf{v})$ that consists of a covector $a$ and a vector $\mathbf{v}$ to the number $\alpha(\mathbf{v})$. If one looks at the corresponding point $[\mathbf{v}]$ and hyperplane $[\alpha]$ in $\mathbb{R}P^n$ then they will incident iff $\alpha(\mathbf{v}) = 0$; since this condition is homogeneous, it will be true for any representatives $\alpha$ and $\mathbf{v}$ for the lines $[\alpha]$ and $[\mathbf{v}]$, respectively.

The projective-geometric basis for *Poincaré duality* is the fact that a $k$-dimensional linear subspace of $\mathbb{R}^{n+1}$ can either be spanned by $k$ linearly-independent vectors in $\mathbb{R}^{n+1}$ or simultaneously annihilated by $n-k+1$ linearly-independent covectors in $\mathbb{R}^{n+1*}$. Thus, the duality is between linear subspaces of complementary dimensions in the two vector spaces $\mathbb{R}^{n+1}$ and $\mathbb{R}^{n+1*}$. The relationship then projects to subspaces of complementary dimension in $\mathbb{R}P^n$ and $\mathbb{R}P^{n*}$, except that the linear combination of vectors gets replaced with the join of points and the simultaneous annihilation by covectors corresponds to the meet of hyperplanes.

Note that although the association of any $k$-dimensional subspace of $\mathbb{R}P^n$ with a unique $n-k-1$-dimensional subspace of $\mathbb{R}P^{n*}$ says nothing about associating points in one projective space with points in the other. Such an association would be more open-ended, and must be defined explicitly.

A *correlation* is an invertible map $[C]: \mathbb{R}P^n \to \mathbb{R}P^{n*}$ such that:

1. Inclusion is inverted: $A \subset B$ implies that $[C](A) \supset [C](B)$
2. The image of a join is a meet: $[C](A \vee B) = [C](A) \vee [C](B)$
3. The image of a meet is a join: $[C](A \cap B) = [C](A) \cap [C](B)$

Thus, points will go to hyperplanes, and the line that is generated by the join of two distinct points will correspond to the meet of the two correlated hyperplanes. That meet will be a line iff $n = 3$. One also sees that since incidence is a symmetrization of inclusion, a correlation will also preserve incidence.

A correlation $[C]$: $\mathbb{RP}^n \to \mathbb{RP}^{n*}$ will be covered by a linear isomorphism $C : \mathbb{R}^{n+1} \to \mathbb{R}^{n+1*}$ that is unique, up to a non-zero scalar multiple. Thus, when a basis has been chosen for $\mathbb{R}^{n+1}$, and its reciprocal basis is chosen for $\mathbb{R}^{n+1*}$, a correlation can be expressed as the system of linear equations:

$$\rho\, y_i = C_{ij}\, x^j, \tag{2.3}$$

in which the determinant of the matrix $C_{ij}$ is non-vanishing.

When a correlation $[C]$: $\mathbb{RP}^n \to \mathbb{RP}^{n*}$ associates a point $x \in \mathbb{RP}^n$ with a hyperplane $[C](x) \in \mathbb{RP}^{n*}$, the point $x$ will be referred to as a *pole* and the hyperplane $[C](x)$ will be referred to as its *polar,* or *polar hyperplane* under that correlation.

An obvious question to ask of the hyperplane $[C](x)$ is whether the point $x$ is incident on it. If that is the case then one will have:

$$C(\mathbf{x})(\mathbf{x}) = 0 \tag{2.4}$$

for any $C : \mathbb{R}^{n+1} \to \mathbb{R}^{n+1*}$ that covers $[C]$ and any $\mathbf{x}$ that covers $x$.

One can, more generally, define the bilinear functional on $\mathbb{R}^{n+1}$:

$$C(\mathbf{x}, \mathbf{y}) = C(\mathbf{x})(\mathbf{y}), \tag{2.5}$$

and if one has (2.4) then one will call that bilinear functional *involutive.* That situation can come about in two different ways:

$$C(\mathbf{x}, \mathbf{y}) = \pm\, C(\mathbf{y}, \mathbf{x}), \tag{2.6}$$

In the positive case, $[C]$ is called a *polarity*, while in the second case, it is called a *null polarity*. A null polarity can exist only when $n + 1$ is even, since it also defines a symplectic structure on $\mathbb{R}^{n+1}$.

Any correlation – involutive or not – will define a quadric hypersurface in $\mathbb{RP}^n$ by way of all $[\mathbf{x}] \in \mathbb{RP}^n$ such that (2.4) is true. If one polarizes the bilinear functional $C$ into the sum of a symmetric functional and an anti-symmetric one:

$$C(\mathbf{x}, \mathbf{y}) = C_+(\mathbf{x}, \mathbf{y}) + C_-(\mathbf{x}, \mathbf{y}), \tag{2.7}$$

with

$$C_\pm(\mathbf{x}, \mathbf{y}) = \tfrac{1}{2}[C(\mathbf{x}, \mathbf{y}) \pm C(\mathbf{y}, \mathbf{x})], \tag{2.8}$$

then one will see that since $C_-(\mathbf{x}, \mathbf{x}) = 0$ for any $\mathbf{x}$, the quadric of $C$ will be defined by only its symmetric part. The quadric that a correlation defines will then take the homogeneous form:

$$C_+(\mathbf{x}, \mathbf{x}) = 0. \tag{2.9}$$

**3. Line geometry.** The basic focus of line geometry is lines in projective spaces, and typically, the lines in $\mathbb{R}P^3$, in particular. Once again, it is important to understand that a line in a projective space will actually mean a *projective* line; i.e., a one-dimensional manifold that is diffeomorphic to $\mathbb{R}P^1$, which is also diffeomorphic to a circle. Hence, in the vector space that covers the projective space a line will be represented by a plane through the origin, so a point of intersection of two lines (if there is one) in a projective space will correspond to a line of intersection of two planes in the vector space that covers it. In particular, two lines in the projective plane will be covered by two planes through the origin of $\mathbb{R}^3$, which will then have to intersect non-trivially; i.e., all lines in the projective plane must intersect (if only at infinity).

*a. The description of lines.* Just as two distinct points *a*, *b* in an affine space will determine a unique line [*a*, *b*], the same can be said in projective space. (Of course, a given line will not define a unique pair of distinct points, but an infinitude of them.) However, since there are two types of points in projective space, there will appear to be three types of lines to begin with. Naively, we define:

1. *Finite lines*: lines for which *a* and *b* are both finite.
2. *Lines at infinity:* lines for which *a* and *b* are both at infinity.
3. *Lines to infinity:* lines for which one point is finite and the other one is infinite.

However, we can see that the finite lines are really lines to infinity due to the fact that:

**Theorem:**

*Any line in a projective space must contain at least one point at infinity.*

**Proof:** Any line in a projective space $\mathbb{R}P^n$ is covered by a plane through the origin in the vector space $\mathbb{R}^{n+1}$ that projects onto $\mathbb{R}P^n$, and thus a two-dimensional linear subspace. The points at infinity in $\mathbb{R}P^n$ are covered by a hyperplane in $\mathbb{R}^{n+1}$; i.e., a linear subspace of codimension one, which is less than two. Thus, any plane through the origin in $\mathbb{R}^{n+1}$ will have to intersect the hyperplane at infinity in a linear subspace of dimension at least one, which will then project to at least one point at infinity in $\mathbb{R}P^n$.

More generally, a subspace of $\mathbb{R}\mathrm{P}^n$ of dimension $k > 0$ must intersect $\mathbb{R}\mathrm{P}^{n-1}_\infty$ in a subspace of dimension $> k - 1$. For instance, any plane in $\mathbb{R}\mathrm{P}^3$ that is not completely at infinity will define a line at infinity.

From the last theorem, we then see that there are only two types of lines in any projective space, namely lines at infinity and lines to infinity. Furthermore, in some treatments of line geometry the lines to infinity are referred to simply as points at infinity, which is reasonable when one confines one's attention to the geometry of points at infinity.

In any event, two distinct points $a$ and $b$ in $\mathbb{R}\mathrm{P}^3$ will be uniquely covered by two distinct lines through the origin [**a**] and [**b**], resp., in $\mathbb{R}^4$, which will then span a unique plane [**a**, **b**] through the origin. A different choice of points $a$, $b$ for generating the line [$a$, $b$] would have the effect of defining a different pair of distinct lines [**a**], [**b**], although one can see that they would still lie in the previous plane. Thus, the association of lines in $\mathbb{R}\mathrm{P}^3$ with planes through the origin in $\mathbb{R}^4$ is one-to-one.

*b. The Plücker-Klein embedding.* The distinct lines [**a**], [**b**] will be (non-uniquely) generated by two non-collinear vectors **a** and **b**. Thus, if one takes the exterior product $\mathbf{a} \wedge \mathbf{b}$ then the result be a non-zero bivector. Since any other pair of non-collinear vectors $\mathbf{a}'$, $\mathbf{b}'$ that lie in the plane [**a**, **b**] will produce the same line [$a$, $b$] in $\mathbb{R}\mathrm{P}^3$, and the two 2-frames for [**a**, **b**] are related by:

$$\mathbf{a}' = A^1_1\mathbf{a} + A^1_2\mathbf{b}, \qquad \mathbf{b}' = A^2_1\mathbf{a} + A^2_2\mathbf{b}, \tag{3.1}$$

with $\det A \neq 0$, one will see that:

$$\mathbf{a}' \wedge \mathbf{b}' = \det A\ \mathbf{a} \wedge \mathbf{b}. \tag{3.2}$$

Thus, any differing choices of 2-frame for the plane [**a**, **b**] will give bivectors that differ only by a non-zero scalar multiple. Hence, under the projection ([1]) $\Lambda_2 - 0 \to P\Lambda_2$, $\mathbf{a} \wedge \mathbf{b} \mapsto [\mathbf{a} \wedge \mathbf{b}]$, the plane that is spanned by any pair **a**, **b** of non-collinear vectors that lie in it will correspond to a unique point $[\mathbf{a} \wedge \mathbf{b}] \in P\Lambda_2$. Putting the two correspondences together, one gets that any line [$a$, $b$] in $\mathbb{R}\mathrm{P}^3$ will define a unique point in $P\Lambda_2$. If one denotes the manifold of lines in $\mathbb{R}\mathrm{P}^3$ by $\mathcal{L}(3)$ then the map $\mathcal{L}(3) \to P\Lambda_2$, $[a, b] \mapsto [\mathbf{a} \wedge \mathbf{b}]$ will be an embedding that is not a surjection, as is easily seen by the fact that not all bivectors in $\Lambda_2$ are decomposable. Indeed, the most general element of $\Lambda_2$ can be given the form:

$$\mathbf{B} = \mathbf{a} \wedge \mathbf{b} + \mathbf{c} \wedge \mathbf{d}, \tag{3.3}$$

---

([1]) Since we shall be concerned with lines in $\mathbb{R}\mathrm{P}^3$ exclusively from now on, we shall abbreviate the notations $\Lambda^4_2$ and $\Lambda^2_4$ by $\Lambda_2$ and $\Lambda^2$, respectively.

in which the vectors **a**, **b**, **c**, **d** are linearly-independent, and thus define a frame for $\mathbb{R}^4$. As it turns out, there are only three types of bivectors over $\mathbb{R}^4$, namely: **B** = 0, **B** decomposable, **B** as in (3.3). If one takes the exterior product **B** ^ **B** then one will see that it is zero for any decomposable bivector and non-zero for any bivector of the form (3.3), namely, **a** ^ **b** ^ **c** ^ **d**. However, **B** ^ **B** ^ **B** = 0 in any case, since all hexa-vectors on a four-dimensional vector space must vanish.

A useful relationship between exterior products and intersections of lines is the following:

**Theorem:**

*If the lines l, l′* $\in \mathcal{L}(3)$ *are represented by decomposable bivectors* **B**, **B′** $\in \Lambda_2$ *then l and l′ intersect non-vacuously iff:*

$$\mathbf{B} \wedge \mathbf{B}' = 0. \tag{3.4}$$

Note that non-vacuous intersection is a weaker constraint than incidence.

If one calls the minimum number of linearly-independent vectors that it takes to represent a bivector **B** its *rank* then one can sees that the bivectors **B** = 0, **a** ^ **b**, **a** ^ **b** + **c** ^ **d** will then have rank 0, 2, 4, respectively. If the lowest exterior power of **B** that vanishes is $p$ then the rank of **B** will be $2(p - 1)$.

Now, the equation:

$$\mathbf{B} \wedge \mathbf{B} = 0 \tag{3.5}$$

is a homogeneous, quadratic equation for the elements of the six-dimensional vector space $\Lambda_2$, so the bivectors that satisfy that equation will define a hypersurface in $\Lambda_2$, which will then have dimension five. Since the equation is homogeneous, it will also define a quadratic equation on the points [**B**] in the (five-dimensional) projective space $\mathrm{P}\Lambda_2$, and thus a four-dimensional projective hypersurface $\mathcal{K}$. One calls this hypersurface the *Klein quadric*. Since every decomposable bivector will define a line in $\mathbb{R}\mathrm{P}^3$, the Klein quadric will be the image of the Plücker-Klein embedding. In particular, the set $\mathcal{L}(3)$ of all lines in $\mathbb{R}\mathrm{P}^3$ will be diffeomorphic to $\mathcal{K}$, and will thus be four-dimensional.

As long as one is dealing with $\mathbb{R}\mathrm{P}^3$, in particular, one can also embed $\mathcal{L}(3)$ in $\mathrm{P}\Lambda^2 \cong (\mathrm{P}\Lambda_2)^*$ with an image that equals the dual $\mathcal{K}^*$ of the Klein quadric, which is covered in $\Lambda^2$ by the set of all decomposable 2-forms $\alpha \in \Lambda^2$, which will then satisfy the quadratic equation:

$$\alpha \wedge \alpha = 0. \tag{3.6}$$

The fact that this embedding works only for a three-dimensional projective space is due to the fact that only then will the meet of two distinct planes always be a line. Thus, a line in $\mathbb{R}\mathrm{P}^3$ gets associated with a (non-unique) pair of distinct 1-forms $\alpha, \beta \in \Lambda^1$, and thus, with a non-zero 2-form $\alpha \wedge \beta$; the original line is then covered in $\mathbb{R}^{n+1}$ by the annihilating plane of $\alpha \wedge \beta$. Any other choice of $\alpha, \beta$ that gives the same line will give a 2-form that differs from $\alpha \wedge \beta$ by a non-zero scalar multiple. Thus, one will get an embedding of $\mathcal{L}(3)$ in $\mathrm{P}\Lambda^2$, as well as in $\mathrm{P}\Lambda_2$ .

Typically, one refers to the six independent components $p_{ij} = x_i\, y_j - x_j\, y_i$ , $i, j = 0, 1, 2, 3$ of a decomposable bivector $\mathbf{x} \wedge \mathbf{y}$ with respect to some basis on $\mathbb{R}^4$ as the "Plückerian coordinates" of the line $[x, y]$ in $\mathbb{R}\mathrm{P}^3$. These six homogeneous coordinates reduce to five inhomogeneous ones for $\mathrm{P}\Lambda_2^4$, and four independent inhomogeneous coordinates when one imposes the constraint that they must lie on the Klein quadric.

Actually, in Plücker's ground-breaking book *Neue Geometrie des Raumes* [**2**] ("New Geometry of Space"), he defined the four coordinates of a line in $\mathbb{R}\mathrm{P}^3$ directly by means of two equations that involved the inhomogeneous coordinates *X, Y, Z,* namely, the equations for the projections of the line onto the *XZ* and *YZ* planes:

$$X = rZ + \rho, \qquad Y = sZ + \sigma. \tag{3.7}$$

(The equation for the *XY* projection can then be derived from these.) The inhomogeneous coordinates of the line are then $(r, s, \rho, \sigma)$.

*c. Time-space decompositions of* $\Lambda_2$ and $\Lambda^2$. If we now introduce the distinction between finite points and points at infinity, as described by a choice of hyperplane through the origin in $\mathbb{R}^4$ that will give a direct-sum decomposition $\mathbb{R}^4 = \mathbb{R} \oplus \mathbb{R}^3$, as above, then that will give a corresponding direct-sum decomposition $\Lambda_2^t \oplus \Lambda_2^s$ of $\Lambda_2$ into a pair of three-dimensional subspaces $\Lambda_2^t$ and $\Lambda_2^s$, where the *t* and *s* suggest "temporal" and "spatial," which will be the case for Minkowski space-time.

By looking at the linear functionals on the direct summands of $\Lambda_2$, one will get a corresponding direct sum decomposition of $\Lambda^2$ into $\Lambda_t^2 \oplus \Lambda_s^2$. Thus, $\Lambda_t^2 = (\Lambda_2^t)^*$ and $\Lambda_s^2 = (\Lambda_2^s)^*$.

Because $\mathbf{B} \wedge \mathbf{B} = 0$ for any bivector $\mathbf{B}$ over a three-dimensional vector space, all bivectors over a three-dimensional space will be decomposable. In particular, the elements of $\Lambda_2^t$ will take the form $\mathbf{t} \wedge \mathbf{a}$ for some $\mathbf{a} \in \mathbb{R}^3$ and those of $\Lambda_2^s$ will take the form of $\mathbf{b} \wedge \mathbf{c}$, where both $\mathbf{b}$ and $\mathbf{c}$ belong to $\mathbb{R}^3$. Hence, a typical element $\mathbf{B} \in \Lambda_2$ will take the form:

$$\mathbf{B} = \mathbf{t} \wedge \mathbf{a} + \mathbf{b} \wedge \mathbf{c}\,. \tag{3.8}$$

If we interpret $\mathbb{R}^3$ as the hyperplane at infinity in $\mathbb{R}^4$ then we will see that the bivectors in $\Lambda_2^t$ correspond to lines of the form $[O, a]$ – i.e., lines to infinity or points at

infinity – and the bivectors in $\Lambda_2^s$ will correspond to lines of the form [$b$, $c$] – i.e., lines at infinity.

As for the case of finite lines, if $\mathbf{a} = a^0\mathbf{t} + \mathbf{a}_s$, $\mathbf{b} = b^0\mathbf{t} + \mathbf{b}_s$, with $a^0$ and $b^0$ non-zero, then:

$$\mathbf{a} \wedge \mathbf{b} = \mathbf{t} \wedge (a^0\, \mathbf{b}_s - b^0\, \mathbf{a}_s) + \mathbf{a}_s \wedge \mathbf{b}_s\,, \tag{3.9}$$

which is then of the same form as (3.8).

In the special case where $a^0 = b^0 = 1$, (3.9) will take the form:

$$\mathbf{a} \wedge \mathbf{b} = \mathbf{t} \wedge (\mathbf{b}_s - \mathbf{a}_s) + \mathbf{a}_s \wedge \mathbf{b}_s\,. \tag{3.10}$$

One also finds this way of representing a line in $\mathbb{R}P^3$ used in Plücker, at least implicitly.

If we define $\mathbf{B} = \mathbf{t} \wedge \mathbf{a} + \mathbf{b} \wedge \mathbf{c}$, more generally, and look for the condition for $\mathbf{B}$ to represent a line – viz., $\mathbf{B} \wedge \mathbf{B} = 0$ – then we will get:

$$\mathbf{B} \wedge \mathbf{B} = \mathbf{t} \wedge \mathbf{a} \wedge \mathbf{b} \wedge \mathbf{c}, \tag{3.11}$$

so since $\mathbf{t}$ is always independent of $\mathbf{a}$, $\mathbf{b}$, $\mathbf{c}$, $\mathbf{B} \wedge \mathbf{B}$ will vanish iff $\mathbf{a} \wedge \mathbf{b} \wedge \mathbf{c}$ does, which says that the vector $\mathbf{a}$ must lie in the plane of $\mathbf{b}$ and $\mathbf{c}$, or that the point [$\mathbf{a}$] must lie along the line [$b$, $c$], where $b$ = [$\mathbf{b}$], $c$ = [$\mathbf{c}$]. We can illustrate this situation as in Fig. 1. In it, we also indicate the more general situation for a bivector $\mathbf{t} \wedge \mathbf{a}' + \mathbf{b} \wedge \mathbf{c}$ that is not decomposable; i.e., it has rank four. The point $a'$ = [$\mathbf{a}$] will not lie on the line [$b$, $c$], in that case.

Note that since $\mathbf{a} \wedge \mathbf{b} \wedge \mathbf{c} \wedge \mathbf{d} = 0$ when all of the vectors involved project to points at infinity, one will also see that any two lines at infinity must intersect. Of course, this is a well-known property of the projective plane, which has finite and infinite points of its own. In particular, lines that are parallel in the affine space that represents the complement of the (projective) line at infinity will intersect at points at infinity, such as the vanishing points for the edges of a highway as one drives into the horizon.

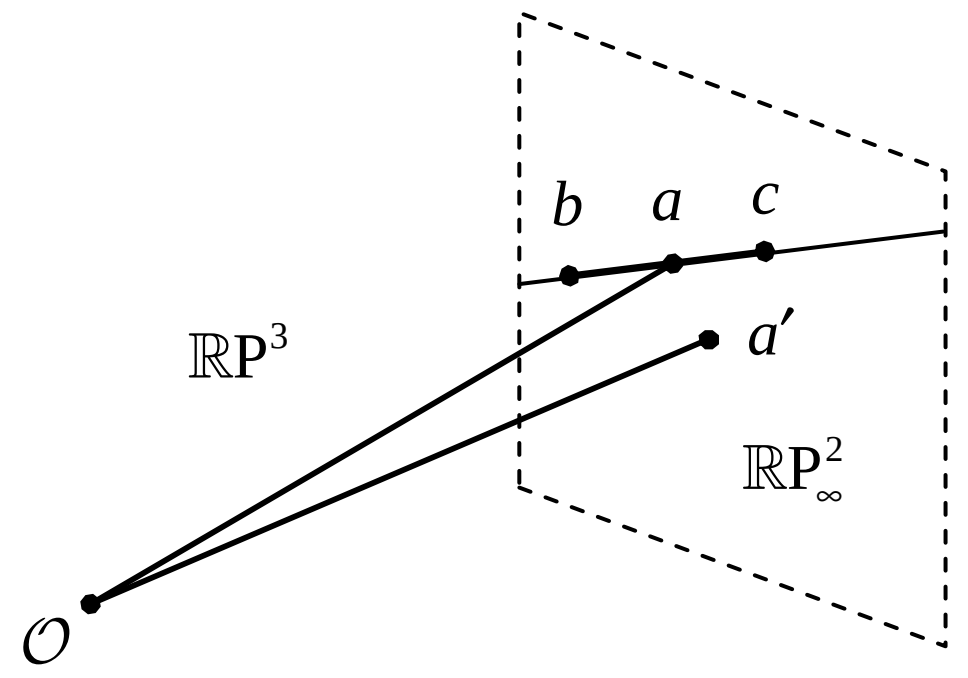

Figure 1.

The Klein quadric can be related to a scalar product on $\Lambda_2$ that is defined by:

$$\langle \mathbf{A}, \mathbf{B} \rangle = V(\mathbf{A} \wedge \mathbf{B}), \tag{3.12}$$

where $V$ is a volume element on $\mathbb{R}^4$; i.e., a non-zero 4-form.

One can also regard this scalar product as being the polarity that is defined by the correlation that takes the form of the Poincaré duality that was mentioned above, which is the correlation [#] : $P\Lambda_2 = \to P\Lambda^2$ that is covered by the linear isomorphism #: $\Lambda_2 \to \Lambda^2$ that takes any bivector **B** to the 2-form:

$$\#\mathbf{B} = i_{\mathbf{B}}V, \tag{3.13}$$

in which $i_{\mathbf{B}} : \Lambda^4 \to \Lambda^2$ is the interior product operator that is defined by **B**.

Hence, one can say that:

$$<\mathbf{A}, \mathbf{B}> = \#\mathbf{A}(\mathbf{B}). \tag{3.14}$$

The quadric that the correlation [#] (which is then a polarity, due to the symmetry of its bilinear functional) defines is the Klein quadric.

When the bivectors **A** and **B** are decomposable, so they represent lines $a$ and $b$ in $\mathbb{R}P^3$, the geometric interpretation of the vanishing of <**A**, **B**> will be that the corresponding lines $a$ and $b$ must intersect. In particular, as we saw above, <**B**, **B**> = 0 iff the line to infinity that it includes intersects the line at infinity, as in Fig. 1.

If $\{\mathbf{e}_\mu, \mu = 0, 1, 2, 3\}$ is a linear frame on $\mathbb{R}^4$ and $\{\theta^\mu, \mu = 0, 1, 2, 3\}$ is its reciprocal coframe (i.e., $\theta^\mu(\mathbf{e}_\nu) = \delta^\mu_\nu$) then one can define a volume element for $\mathbb{R}^4$ by means of:

$$V = \theta^0 \wedge \theta^1 \wedge \theta^2 \wedge \theta^3 = \frac{1}{4!}\varepsilon_{\mu_0\mu_1\mu_2\mu_3}\theta^{\mu_0} \wedge \theta^{\mu_1} \wedge \theta^{\mu_2} \wedge \theta^{\mu_3} \tag{3.15}$$

and a linear frame $\{\mathbf{E}_\mathrm{I}, \mathrm{I} = 1, \ldots, 6\}$ for $\Lambda_2$ by way of:

$$\mathbf{E}_i = \mathbf{e}_0 \wedge \mathbf{e}_i\ , \quad \mathbf{E}_{i+3} = \tfrac{1}{2}\varepsilon_{ijk}\,\mathbf{e}_j \wedge \mathbf{e}_k \qquad (i = 1, 2, 3). \tag{3.16}$$

Thus, if $\mathbf{e}_0$ is assumed to project to a finite point, while $\mathbf{e}_1$, $\mathbf{e}_2$, $\mathbf{e}_3$ project to points at infinity then the first three basis vectors $\mathbf{E}_i$ will span a three-dimensional subspace that represents lines to infinity, while the last three $\mathbf{E}_{i+3}$ span a three-dimensional subspace that represents the lines at infinity.

One then finds that:

$$<\mathbf{E}_i, \mathbf{E}_j> = <\mathbf{E}_{i+3}, \mathbf{E}_{i+3}> = 0, \qquad <\mathbf{E}_i, \mathbf{E}_{j+3}> = <\mathbf{E}_{i+3}, \mathbf{E}_j> = \delta_{ij}, \tag{3.17}$$

so this basis is not orthogonal for the scalar product, and its matrix with respect to this linear frame will be, in block form:

$$V_{\mathrm{IJ}} = \left[\begin{array}{c|c} 0 & \delta_{ij} \\ \hline \delta_{ij} & 0 \end{array}\right]. \tag{3.18}$$

In order to find an orthonormal basis for $\Lambda_2$, one introduces the linear isomorphism * of $\Lambda_2$ with itself that makes:

$$*\mathbf{E}_i = \mathbf{E}_{i+3}, \qquad *\mathbf{E}_{i+3} = -\mathbf{E}_i, \tag{3.19}$$

so:

$$*^2 = -I. \tag{3.20}$$

One can then define the linear frame:

$$\bar{\mathbf{E}}_i = \tfrac{1}{2}(\mathbf{E}_i + *\mathbf{E}_i), \qquad \bar{\mathbf{E}}_{i+3} = \tfrac{1}{2}(\mathbf{E}_i - *\mathbf{E}_i), \tag{3.21}$$

which is orthogonal, since:

$$<\bar{\mathbf{E}}_i, \bar{\mathbf{E}}_j> = -<\bar{\mathbf{E}}_{i+3}, \bar{\mathbf{E}}_{j+3}> = \delta_{ij}, \qquad <\bar{\mathbf{E}}_i, \bar{\mathbf{E}}_{j+3}> = <\bar{\mathbf{E}}_{i+3}, \bar{\mathbf{E}}_j> = 0. \tag{3.22}$$

Hence, the matrix for the scalar product relative to this frame will be:

$$\bar{V}_{IJ} = \left[\begin{array}{c|c} \delta_{ij} & 0 \\ \hline 0 & -\delta_{ij} \end{array}\right]. \tag{3.23}$$

A further property of the orthonormal basis is that:

$$*\bar{\mathbf{E}}_i = -\bar{\mathbf{E}}_{i+3}, \quad *\bar{\mathbf{E}}_{i+3} = \bar{\mathbf{E}}_i. \tag{3.24}$$

*d. Line complexes.* A *line complex* [**2, 9-11**] in $\mathbb{R}P^3$ is a zero-locus of some function $F: \mathcal{L}(3) \to \mathbb{R}$, so it can also be represented as $F: \mathcal{K} \to \mathbb{R}$, $[\mathbf{a} \wedge \mathbf{b}] \mapsto F[\mathbf{a} \wedge \mathbf{b}]$; thus, it will define some set of lines in $\mathbb{R}P^3$. Since $\mathcal{K}$ is a quadric hypersurface in $\Lambda_2$, which is then defined by a homogeneous quadratic equation on $\Lambda_2$, one can typically represent a line complex in $\mathbb{R}P^3$ by a pair of homogeneous equations on $\Lambda_2$:

$$f(\mathbf{B}) = 0, \qquad \mathbf{B} \wedge \mathbf{B} = 0, \tag{3.25}$$

in which $f: \Lambda_2 \to \mathbb{R}$ agrees with $F[\mathbf{a} \wedge \mathbf{b}]$, up to a non-zero scalar multiple. Thus, the line complex that one defines becomes the intersection of two hypersurfaces in $\Lambda_2$, one of which will always be the Klein quadric.

In the classical literature, the function $f$ was usually an elementary algebraic function, such as a linear or quadratic function. In the former case, one then referred to a *linear line complex* – or simply, a *linear complex* – and in the latter case, to a *quadratic line complex.*

Since a linear complex is defined by a linear functional on $\Lambda_2$, one can also think of it as defined by a 2-form $\Omega \in \Lambda^2 \cong (\Lambda_2)^*$. The action of $\Omega$ on a decomposable bivector $\mathbf{a} \wedge \mathbf{b}$ is:

$$\Omega(\mathbf{a} \wedge \mathbf{b}) = \Omega(\mathbf{a}, \mathbf{b}), \tag{3.26}$$

where the right-hand side refers to the definition of $\Omega$ as an anti-symmetric bilinear functional on vectors in $\mathbb{R}^4$. One can then extend $\Omega$ to non-decomposable bivectors by demanding that it be linear, although we will be primarily concerned with the decomposable ones, anyway.

If one introduces the Poincaré isomorphisms $\#: \Lambda_k \to \Lambda^{4-k}$, $\mathbf{A} \mapsto i_{\mathbf{A}}V$ that are defined by a choice of volume element $V$ on $\mathbb{R}^4$ then $\#: \Lambda_2 \cong \Lambda^2$, so one can also make $\Omega = \#\mathbf{A}$ for a unique bivector $\mathbf{A}$, and the action of $\Omega$ on any other bivector $\mathbf{B}$ will be:

$$\Omega(\mathbf{B}) = \#\mathbf{A}(\mathbf{B}) = V(\mathbf{A} \wedge \mathbf{B}) = <\mathbf{A}, \mathbf{B}>. \tag{3.27}$$

Thus, the vanishing of $\Omega(\mathbf{B})$ is equivalent to the vanishing of $\mathbf{A} \wedge \mathbf{B}$, or, when both bivectors are decomposable, the intersection of the lines that they represent.

The situation in which $\mathbf{A}$ is decomposable (and thus, $\Omega = \#\mathbf{A}$, as well), and thus represents a line in $\mathbb{RP}^3$, defines what one calls a *special linear complex*, and the line that $\mathbf{A}$ corresponds to is called the *axis* of the complex. In such a case, one can then characterize all lines in $\mathbb{RP}^3$ whose corresponding bivector $\mathbf{B}$ makes $\Omega(\mathbf{B})$ vanish by the fact that they will represent all of the lines that intersect the axis of the special linear complex in question.

More generally, when $\Omega$ is not decomposable, it will be represented by two non-intersecting lines, one of which is at infinity and the other of which goes to infinity. If $\mathbf{B}$ represents a line then the vanishing of $\Omega(\mathbf{B})$ will say that the line of $\mathbf{B}$ must intersect both of the lines that correspond to $\Omega$.

Since # takes decomposable bivectors to decomposable 2-forms (i.e., lines to lines), the condition on $\mathbf{A}$ that makes it a line – viz., $\mathbf{A} \wedge \mathbf{A} = 0$ – will correspond to the condition $\Omega \wedge \Omega = 0$ on $\Omega$, which will make the linear complex that $\Omega$ defines special, and which means that $\Omega$ will be (non-uniquely) representable in the form $\alpha \wedge \beta$ for some appropriate 1-forms $\alpha$ and $\beta$. Thus, the special linear complexes will all lie on a quadric $\Omega \wedge \Omega = 0$ in $\Lambda^2$, which will project to the $\mathcal{K}^*$ quadric in $P\Lambda^2$ that will be diffeomorphic to the Klein quadric, since the vector spaces $\Lambda_2$ and $\Lambda^2$ are linearly isomorphic.

One can use the coframe $E^I$ that is reciprocal to a chosen frame $\mathbf{E}_I$ on $\Lambda_2$ to define a linear isomorphism $\Lambda^2 \cong \mathbb{R}^{6*}$, $\Omega \mapsto \Omega^I$. The action of $\Omega$ on $\mathbf{A}$ will then correspond to the linear combination $\Omega^I A_I$.

In the case where $\Omega$ is not decomposable, one also finds that the linear map $\mathbb{R}^4 \to \mathbb{R}^{4*}$ that takes any vector $\mathbf{v}$ to the 1-form $\mathbf{v}^* = i_{\mathbf{v}}\Omega$ will be an isomorphism. Hence, $\Omega$ will define a non-degenerate, anti-symmetric bilinear form on $\mathbb{R}^4$; i.e., a symplectic structure.

This linear isomorphism will correspond to a correlation $\mathbb{R}P^3 \to \mathbb{R}P^{3*}$. If one then thinks of the linear complex $\Omega$ as something that also takes points to planes, as well as lines to lines, then since:

$$\mathbf{v}^*(\mathbf{v}) = \Omega(\mathbf{v}, \mathbf{v}) = 0, \tag{3.28}$$

one can say that the point [**v**] in $\mathbb{R}P^3$ that is represented by the vector **v** must be incident with the plane [$\mathbf{v}^*$] in $\mathbb{R}P^3$ that it is dual to. The correlation that is defined by $\Omega$ is then a null polarity.

A *quadratic line complex* is defined by a quadratic form $Q(\mathbf{A})$ on $\Lambda_2$, which is then related to a bilinear form $Q(\mathbf{A}, \mathbf{B})$ on $\Lambda_2$ . Relative to a chosen frame $\mathbf{E}_I$ on $\Lambda_2$, one can then regard $Q$ as a quadratic form on $\mathbb{R}^6$:

$$Q(\mathbf{A}, \mathbf{B}) = Q_{IJ} A^I B^J, \tag{3.29}$$

in which $Q_{IJ}$ is symmetric, but not necessarily non-degenerate. Thus, if $\mathbf{E}_I$ is an orthonormal frame for $Q$ then the diagonal might contain zeroes, in addition to ± 1's.

The vanishing of $Q(\mathbf{A})$ will define a quadric on $\Lambda_2$, and by projection, on $P\Lambda_2$, as well. Hence, the lines in $\mathbb{R}P^3$ that belong to the quadratic line complex that $Q$ defines will be represented by points that lie in the intersection of the Klein quadric with the quadric that $Q$ defines.

The equations of any quadratic line complex will then take the homogeneous form:

$$\mathbf{B} \wedge \mathbf{B} = 0, \qquad Q(\mathbf{B}) = 0. \tag{3.30}$$

**4. Applications to mechanics.** The application of line geometry to mechanics is already rather voluminous (cf., e.g., [**12-14**]), due to the work of Nineteenth-Century geometers such as Poinsot, Chasles, Plücker, Klein, Grassmann, Sir Robert Ball, Eduard Study, and a host of others. It continues to be applied to mechanics to this day, but mostly by mechanical engineers who deal in structures and mechanisms (cf., e.g., [**15-17**]. Therefore, we shall only mention those aspects of the mechanical applications that might be analogous to corresponding applications to electromagnetism.

The ones that will be of interest to us are initially based in the fact that in the theory of mechanical moments the six-dimensional vector spaces $\Lambda_2$ and $\Lambda^2$ appear naturally as a space for the representation of the six-dimensional Lie algebra $\mathfrak{iso}(3)$ of infinitesimal rigid motions in three-dimensional Euclidian space and its dual vector space $\mathfrak{iso}(3)^*$.

*a. The representation of infinitesimal rigid motions.* Although historically the work of Michel Chasles came after that of Poinsot, nonetheless, since Chasles was addressing kinematics and Poinsot was addressing kinetics or dynamics, we will start with Chasles's theorem.

What Chasles proved was that when any rigid motion $T \in ISO(3)$ acted on three-dimensional Euclidian space $E^3$ (which we regard as an affine space $A^3$ on whose tangent spaces a Euclidian metric has been defined) there would be a unique *central axis* for the motion; i.e., a line $l$ in $E^3$ such that the rigid motion could be given the canonical form in which it consisted of a translation along the axis and a rotation about it; that is, a rotation in any plane that was perpendicular to the axis.

In order to relate this to $\Lambda_2$, one then goes to infinitesimal generators of rigid motions – i.e., elements of the Lie algebra $\mathfrak{iso}(3)$. As a vector space, it will decompose into a direct sum $\mathbb{R}^3 \oplus \mathfrak{so}(3)$ of two three-dimensional subspaces, although as a Lie algebra, this will be a semi-direct sum. The $\mathbb{R}^3$ subalgebra represents the infinitesimal generators of one-parameter subgroups of translations, while the $\mathfrak{so}(3)$ sub-algebra will represent the infinitesimal generators of one-parameter subgroups of rotations.

Although a translation is typically represented by a displacement vector in $\mathbb{R}^3$, an infinitesimal rotation can be represented by either a 3-vector $\boldsymbol{\omega}$ or an anti-symmetric matrix $\omega = \mathrm{ad}(\boldsymbol{\omega})$, where the notation ad refers to the adjoint representation of the Lie algebra $\mathfrak{so}(3)$ – when it is regarded as $\mathbb{R}^3$, given the vector cross product – in the Lie algebra $\mathfrak{L}(3; \mathbb{R})$ of linear maps from $\mathbb{R}^3$ to itself, when it is given the commutator bracket $[A, B] = AB - BA$. Under this representation, the linear map $\mathrm{ad}(\boldsymbol{\omega})$ will take any vector $\mathbf{r} \in \mathbb{R}^3$ to $[\boldsymbol{\omega}, \mathbf{r}] = \boldsymbol{\omega} \times \mathbf{r}$. One then recognizes this as the usual way of defining the tangential velocity to a circular motion about some reference point. Similarly, if $\mathbf{v}$ is that tangential velocity then $\mathbf{v} \times \mathbf{r}$ will give the "orbital" angular velocity $\boldsymbol{\omega}$.

Since the matrix $\omega^i_j$ of $\mathrm{ad}(\boldsymbol{\omega})$ will be anti-symmetric for any choice of basis $\{\mathbf{e}_1, \mathbf{e}_2, \mathbf{e}_3\}$ on $\mathbb{R}^3$, one can raise the lower index to obtain the anti-symmetric matrix $\omega^{ij}$, which can be used as the components of a bivector in $\Lambda_2(\mathbb{R}^3)$:

$$\boldsymbol{\omega} = \tfrac{1}{2}\,\omega^{ij}\,\mathbf{e}_i \wedge \mathbf{e}_j\,. \tag{4.1}$$

Thus, if $\mathbb{R}^3$ is included in $\mathbb{R}^4$ as summand in $\mathbb{R} \oplus \mathbb{R}^3$, and $\mathbb{R}^4$ represents the homogeneous coordinates of $\mathbb{R}P^3$ then $\boldsymbol{\omega}$ will represent a line at infinity.

In order to justify representing an infinitesimal displacement vector $\mathbf{v}$ in the form $\mathbf{e}_0 \wedge \mathbf{v}$, which would then represent a line to infinity, we need only point out that the vector space $\mathbb{R}^3$ is linearly isomorphic to the vector space $\mathbf{e}_0 \wedge \mathbb{R}^3$ by the map associates the basis $\mathbf{e}_i$ for $\mathbb{R}^3$ with the basis $\mathbf{e}_0 \wedge \mathbf{e}_i$ for $\mathbf{e}_0 \wedge \mathbb{R}^3$. Thus, the element $\boldsymbol{\Omega} = \mathbf{v} + \boldsymbol{\omega} \in \mathfrak{iso}(3)$ will correspond to an element

$$\boldsymbol{\Omega} = \mathbf{e}_0 \wedge \mathbf{v} + \boldsymbol{\omega} = \mathbf{e}_0 \wedge v^i\,\mathbf{e}_i + \tfrac{1}{2}\,\omega^{ij}\,\mathbf{e}_i \wedge \mathbf{e}_j \tag{4.2}$$

in $\Lambda_2$ .

Since we are not introducing a Lie bracket on $\Lambda_2$, the difference between $\mathbf{a} \times \mathbf{b}$ and $\mathbf{a} \wedge \mathbf{b}$ becomes inessential.

The condition that Ω must represent a line – namely, that $\mathbf{\Omega} \wedge \mathbf{\Omega} = 0$ – then gives the condition that **v** must lie in the plane of **ω**. Hence, the general element of $\mathfrak{iso}(3)$ will define two non-intersecting lines, one of which (the central axis) goes to infinity, and the other of which is skew and perpendicular to it, and lies at infinity.

One can also think of the elements of $\mathfrak{iso}(3)$ as virtual displacements that take the form $\delta\mathbf{s} + \delta\boldsymbol{\theta}$ of the sum of a virtual translations $\delta\mathbf{s}$ and a virtual rotation $\delta\boldsymbol{\theta}$, which will then correspond to a bivector:

$$\mathbf{\Omega} = \mathbf{e}_0 \wedge \delta x^i \, \mathbf{e}_i + \tfrac{1}{2} \delta\theta^{ij} \, \mathbf{e}_i \wedge \mathbf{e}_j . \tag{4.3}$$

*b. The representation of forces and moments.* What Poinsot had proved previous to Chasles was, in a sense, the dual to Chasles's theorem, namely, that if one had a finite set $\{\mathbf{F}_a, a = 1, \ldots, N\}$ of force vectors at various points $\{\mathbf{r}_a, a = 1, \ldots, N\}$ in a rigid body then the collective effect of the force distribution would be that of a force **F** that acted along a line $l$ that one again calls a central axis and a force moment $\mathbf{M} = \mathbf{r} \times \mathbf{F}$ that acted about that axis – i.e., in any plane that is perpendicular to it.

The sense in which that statement is dual to Chasles's theorem is that one can think of forces and moments as being things that live in the dual space $\mathfrak{iso}(3)^*$ to the Lie algebra $\mathfrak{iso}(3)$, which we think of as containing virtual displacements. One then sees that the effect of evaluating a linear functional $F + M$ in $\mathfrak{iso}(3)^*$ on a virtual displacement $\delta\mathbf{s} + \delta\boldsymbol{\theta}$ in $\mathfrak{iso}(3)$ is to produce a virtual work:

$$\delta W = (F + M)(\delta\mathbf{s} + \delta\boldsymbol{\theta}) = F(\delta\mathbf{s}) + M(\delta\boldsymbol{\theta}) = F_i \, \delta x^i + \tfrac{1}{2} M_{ij} \, \delta\theta^{ij} , \tag{4.4}$$

which takes the form of the corresponding virtual work that is done by the virtual displacement.

The association of $F + M$ with a 2-form on $\mathbb{R}^4$ is then given by associating the reciprocal coframe for $\mathfrak{iso}(3)^*$ with the reciprocal coframe for $\Lambda^2$:

$$L = \theta^0 \wedge F_i \, \theta^i + \tfrac{1}{2} M_{ij} \, \theta^i \wedge \theta^j. \tag{4.5}$$

The condition for $L$ to define a line – namely, $L \wedge L = 0$ – implies that the force $F$ must act in the plane of the moment $M$. In the general case, the force acts along the central axis, which goes to infinity, and the moment defines a line at infinity that is skew and perpendicular to it.

One sees that when the dual object to an infinitesimal rigid motion is represented by a 2-form $L$, it will also define a linear complex on $\mathbb{RP}^3$ that might or might not be special.

It was these dual objects to infinitesimal rigid motions that were originally referred to as "wrenches," "torsors," and "dynames" by the Nineteenth-Century geometers.

*c. Constitutive laws, correlations, and quadratic line complexes.* Commonly, the way that one gets from kinematical states in $\mathfrak{iso}(3)$ to the corresponding dynamical or kinetic states in $\mathfrak{iso}(3)^*$ is by way of a correlation; i.e., a linear isomorphism $C : \mathfrak{iso}(3) \to \mathfrak{iso}(3)^*$, although that generally restricts one to a linear theory, especially, when the dual object represents the force and moment that is produced by an infinitesimal rigid motion. However, when the dual object is the linear and angular momentum that gets associated with a given linear and angular velocity, it is usually more acceptable to treat that as the linear isomorphism that gets associated with a quadratic form on $\mathfrak{iso}(3)$ that one can call the *total kinetic energy density:*

$$E(\Omega) = \tfrac{1}{2}\rho\,\delta_{ij}\,v^i\,v^j + \tfrac{1}{2}I_{ij}\,\omega^i\,\omega^j, \tag{4.6}$$

in which $\rho$ means a mass density and $I_{ij}$ represents a moment of inertia with respect to some chosen axis.

Thus, one usually has a quadratic line complex that gets defined on the lines in $\mathbb{R}\mathrm{P}^3$ that represent infinitesimal rigid motions by way of the linear and angular kinetic energy. In the case of linear forces and torques, one might also have another quadratic line complex that gets defined by Hooke-law-type constitutive laws.

One can also use $\Lambda_2$ to represent the elements of the six-dimensional Lie algebra $\mathfrak{so}(3, 1)$ of infinitesimal Lorentz transformations, in which the infinitesimal translation is replaced with the infinitesimal boost. Indeed, one can think of an infinitesimal translation as essentially a Newtonian approximation to a boost. However, a discussion of relativistic mechanics would take us too far afield at the moment, although one might confer some of the author's observations on the place of projective geometry in special relativity [**18**], as well those of Klein [**19**] and Gschwind [**20**].

*d. Rest spaces in special relativity.* We should, nonetheless, point out that the concept of a "rest space" in special relativity is subtly related to $\mathbb{R}\mathrm{P}^3$, more than it is to $\mathbb{R}^3$. This comes from the fact that the way that one converts relativistic velocity four-vectors, with components $(u^0, u^i)$ to non-relativistic velocity 3-vectors, with components $(v^i)$ involves more than a simple Cartesian (i.e., orthogonal) projection that would simply drop the temporal component $u^0$. Since $u^\mu = dx^\mu / d\tau$, where $\tau$ is the proper-time curve parameter, and $v^i = dx^i / dt$, where $t$ is the time coordinate, one must also convert the curve parameterization in the process of projecting:

$$\frac{dx^i}{dt} = \frac{d\tau}{dt}\frac{dx^i}{d\tau}, \tag{4.7}$$

which makes:

$$v^i = \frac{u^i}{u^0}. \tag{4.8}$$

Thus, the spatial 3-velocity has components that are equal to the *inhomogeneous coordinates* of a point in $\mathbb{R}P^3$, whose homogeneous coordinates are given by the relativistic four-velocity.

Now, the rest space of a motion is characterized by the equality $\tau = t$; i.e., $u^0 = 1$. Thus, velocities, as observed in the rest space will have components of the form $(1, v^i)$, which can be regarded as describing the combination of a point $O$ at infinity that serves as an "origin" for the hyperplane at infinity and a vector that is tangent to $O$ and whose components are $v^i$. This tends to suggest that, at least as far as velocity is concerned, the rest space is a projective space, not an affine one.

Indeed, Max Born once pointed out that all measurements are carried out in the rest space of the measuring devices, while one recalls that projective geometry started out as the geometry of visual perception. It is then intriguing to contemplate the possibility that measuring devices "see" four dimensions in a manner that it is analogous to the way that the human eye "sees" three dimensions; i.e., as if it were a space of homogeneous coordinates for the image that is projected onto the observer/measurer.

**5. Application to electromagnetism.** Now that it is widely known (though perhaps not widely accepted) that Maxwell's theory of electromagnetism can be concisely formulated in terms of exterior differential forms on a four-dimensional differentiable manifold, and especially 2-forms and bivector fields, it becomes immediately clear that the fundamental fields of that theory can be interpreted in terms of line geometry.

*a. Pre-metric electromagnetism.* First, let us briefly review the "pre-metric" formulation of Maxwell's equations ([1]), which was the author's original motivation for exploring the role of projective geometry in physics. The basic question that suggested itself was "What would it mean to be doing 'pre-metric' spacetime geometry?" In the spirit of Klein's Erlanger Programm, the most reasonable answer seemed to be: projective geometry, and more specifically, line geometry.

The basic fields in pre-metric electrodynamics are the usual Minkowski field strength 2-form $F$, the excitation bivector field $\mathfrak{H}$, which describes the response of the medium to the presence of $F$, and the electric current vector field $\mathbf{J}$, which serves as the source of the field $\mathfrak{H}$. One also assumes that the spacetime manifold $M$ is four-dimensional, orientable, and given a specific choice of unit-volume element $V \in \Lambda^4 M$. Thus, one has the Poincaré isomorphisms $\#: \Lambda_k M \to \Lambda^{4-k} M$, $\mathbf{A} \mapsto \#\mathbf{A} = i_{\mathbf{A}} V$, which also allows one to define a generalized divergence operator $\delta: \Lambda_k M \to \Lambda_{k-1} M$, $\mathbf{A} \mapsto \delta\mathbf{A}$ on multi-vector fields, where:

$$\delta = \#^{-1} \cdot d \cdot \#. \tag{5.1}$$

Thus, the generalized divergence operator is the adjoint to the exterior derivative operator $d$ under the Poincaré isomorphisms. It is simple to verify that when $\mathbf{X}$ is a vector field on $M$ that is expressed with respect to a holonomic local frame field, such as a natural frame field $\partial_\mu = \partial / \partial x^\mu$ for a local coordinate system $(U, x^\mu)$, as $\mathbf{X} = X^\mu \partial_\mu$, $\delta\mathbf{X}$ will agree with the usual divergence:

$$\delta\mathbf{X} = \partial_\mu X^\mu. \tag{5.2}$$

([1]) Cf., the author's book [**21**], as well as that of Hehl and Obukhov [**22**], and the references that are cited in them.

The Maxwell equations, in their pre-metric form, are then:

$$dF = 0, \qquad \delta\mathfrak{H} = \mathbf{J}, \qquad \delta\mathbf{J} = 0, \qquad \mathfrak{H} = C(F). \tag{5.3}$$

The first equation is common to the metric formulation of Maxwell's equations. The second one is a slight alteration of the metric form, in that one often uses the codifferential operator $\delta$ that acts on $k$-forms, not $k$-vector fields, so rather than $\mathfrak{H}$, one uses a 2-form $H$, and the vector field $\mathbf{J}$ becomes a 1-form. However, this obscures the fact that the divergence operator is intrinsically related to volume elements, and the introduction of a metric into its definition is basically superfluous.

The third equation in this set is the compatibility constraint on the source field that is derived from the fact that:

$$\delta^2 = \#^{-1} \cdot d \cdot \#\#^{-1} \cdot d \cdot \# = \#^{-1} \cdot d^2 \cdot \# = 0. \tag{5.4}$$

Thus, no vector field $\mathbf{J}$ that does not have vanishing divergence can be the source of the electromagnetic excitation bivector field.

The final equation in the set is the electromagnetic constitutive law for the medium in which the fields exist. It basically amounts to a vector bundle map $C : \Lambda^2M \to \Lambda_2M$ that restricts to a diffeomorphism on each fiber of the vector bundle $\Lambda^2M$. That would correspond to a constitutive law that is nonlinear and non-dispersive, since more generally, the map might consist of an algebraic, differential, and integral operator combined. In the case of a linear, non-dispersive medium the map $C$ restricts to a linear isomorphism on each fiber, and thus defines a fourth-rank tensor field on $M$:

$$C = \frac{1}{4} C^{\kappa\lambda\mu\nu} \partial_\kappa \wedge \partial_\lambda \otimes \partial_\mu \wedge \partial_\nu . \tag{5.5}$$

Thus, $C$ is anti-symmetric in its first two and last two slots as a quadrilinear functional on covectors. If one thinks of each fiber of $\Lambda^2M$ as a vector space, in its own, right, then $C$ defines a doubly-covariant second-rank tensor field:

$$C = C^{IJ}\, \mathbf{E}_I \otimes \mathbf{E}_J \tag{5.6}$$

with no particular symmetry, and thus a bilinear functional on 2-forms.

*b. Line-geometric interpretation of basic electromagnetic structures.* Now that we have all of the basic fields at our disposal, the relationships to line geometry become straightforward. However, one finds that as long as one is not concerned explicitly with the topology of spacetime, it is entirely sufficient to replace the general four-dimensional manifold $M$ with $\mathbb{R}^4$ for the purposes of projective geometry. More generally, one says that one is simply considering the tangent and cotangent spaces to $M$, along with their projectivizations; i.e., the projective spaces that they project onto as vector spaces, which will then projectively equivalent to $\mathbb{RP}^3$ and $\mathbb{RP}^{3*}$, respectively.

If one recalls the basic definitions of the **E** and **B** fields as forces that act upon unit electric charges and moments that act upon unit magnetic dipoles, resp. ([1]), then one can easily see the interpretation of $F$ as a "dyname" that acts upon unit-charge with a unit dipole. Since $\mathfrak{H}$ is dual to $F$, one can then think of the constituent **D** and **H** fields as, in some sense, virtual displacements, and that would certainly be consistent with the classical terminology, but their relevance to kinematics is more tenuous than the association of **E** and **B** with dynamics.

Both the fields $F$ and $\mathfrak{H}$ can be interpreted directly in terms of line geometry, namely, at each point they will define one or two lines in the tangent projective spaces; the actual number will depend upon the rank of the fields in question. Hence, one should examine the kinds of electromagnetic fields that are represented by 2-forms and bivector fields of both ranks. The main difference between $F$ and $\mathfrak{H}$ in the eyes of line geometry is that $F$ defines a linear complex, while $\mathfrak{H}$ defines a pair of lines in $\mathbb{RP}^3$.

When $F$ has rank two, it will define a special linear complex, and the line that it defines in $\mathbb{RP}^3$ will then be the axis of the complex. Thus, lines in $\mathbb{RP}^3$ will belong to the special complex that is defined by a rank-two $F$ iff they intersect the axis that it defines.

We shall assume that the tangent and cotangent spaces are endowed with a choice of time-space splitting that corresponds to the splitting of local coordinates into one temporal coordinate $x^0$ and three spatial coordinates $x^i$, $i = 1, 2, 3$. We will also regard these spacetime coordinates as homogeneous coordinates for $\mathbb{RP}^3$, while $x^0 = 0$ as the equation for its plane at infinity.

All of the most elementary electromagnetic fields are represented by 2-forms of rank two – i.e., the decomposable ones. Namely, static electric fields will take the forms $F = dx^0 \wedge E$ and $\mathfrak{H} = \partial_0 \wedge \mathbf{D}$, which are then lines at infinity and lines to infinity, resp. Static magnetic fields will take the forms $F = \frac{1}{2}\varepsilon_{ijk} B^i\, dx^j \wedge dx^k$ and $\mathfrak{H} = \frac{1}{2}\varepsilon^{ijk} H_i\, \partial_j \wedge \partial_k$, which are just the opposite kinds of lines. The fact that they have rank two follows from the fact that they are 2-forms and bivector fields on three dimensional spaces, so one sees that they will represent the lines at infinity. A third type of rank-two electromagnetic field is that of the fields of electromagnetic waves, which take the forms $F = ik \wedge \mathcal{E}$ and $\mathfrak{H} = i\mathbf{k} \wedge \mathcal{D}$. This case is more geometrically involved, so we will postpone a more detailed discussion to the next of this pair of articles.

---

([1]) Of course, many authors in electromagnetism prefer to think of **B** as a force that acts upon a unit current, à la the Lorentz force. However, one now sees that the interpretation of **B** as a moment makes the association with line geometry more direct.

*c. Electromagnetic interpretation of the Klein quadric.* If one expresses $F$ in the time-space form:

$$F = dt \wedge E - \#_s\mathbf{B} \qquad (\#_s\mathbf{B} = \tfrac{1}{2}\varepsilon_{ijk}\, B^i\, dx^j \wedge dx^k) \tag{5.7}$$

then:

$$F \wedge F = -\,2\, dt \wedge E \wedge \#_s\mathbf{B} = -\,2\, E(\mathbf{B})\, V. \tag{5.8}$$

Hence, the condition for $F$ to lie on the Klein quadric is that:

$$E(\mathbf{B}) = 0; \tag{5.9}$$

note that it is still not necessary to introduce any sort of metric, although this condition usually gets represented in the Euclidian form $\mathbf{E} \cdot \mathbf{B} = 0$.

One finds, similarly, that the condition for:

$$\mathfrak{H} = \partial_t \wedge \mathbf{D} + \#_s^{-1}(H) \qquad (\#_s^{-1}(H) = \tfrac{1}{2}\varepsilon^{ijk}\, H^i\, \partial_j \wedge \partial_k) \tag{5.10}$$

to lie on the corresponding Klein quadric is that:

$$H(\mathbf{D}) = 0 \qquad (\text{i.e., } \mathbf{H} \cdot \mathbf{D} = 0). \tag{5.11}$$

*d. Constitutive laws, correlations, and quadratic line complexes.* Since $F$ is a 2-form, it will define a linear complex in each tangent $\mathbb{R}P^3$ by way of all the bivectors that it annihilates at each point. When $F$ represents an elementary field that would make it decomposable, the linear complex will be special, and its axis will be the line that $F$ defines. A natural question to ask is: What physical significance can one ascribe to the function:

$$F(\mathfrak{H}) = F(C(F)) = C_s(F, F), \tag{5.12}$$

in particular? In this, we are letting $C_s$ denote the symmetric part of the tensor field $C$:

$$C_s(A, B) = \tfrac{1}{2}[C(A, B) + C(B, A)]. \tag{5.13}$$

Thus, when a fiber of $\Lambda^2 M$ is regarded as a six-dimensional vector space by the linear isomorphism $E_{\text{I}} : \Lambda_x^2 M \to \mathbb{R}^6$, $A \mapsto A_{\text{I}} = A(E_{\text{I}})$, the symmetric part of a (linear, non-dispersive) constitutive law will define a quadratic form on $\Lambda_x^2 M$ (or $\mathbb{R}^6$) that can be represented by:

$$C_s(A) = C^{\text{IJ}} A_{\text{I}} A_{\text{J}}\,. \tag{5.14}$$

Hence, such a constitutive law will be associated with a quadratic line complex on $\mathbb{R}P^3$.

When one expands (5.14) using $F$ in the form (5.7), one will get:

$$C_s(F) = C^{(ij)} E_i E_j - 2\, C^{(ij)} E_i B_j + C_{(i+3,\, j+3)} B^i B^j. \tag{5.15}$$

The coupling terms between electric and magnetic field strengths in a constitutive law are often associated with Faraday rotation and optical activity. However, these effects generally contribute to $C_{IJ}$ in an anti-symmetric way, so their symmetric parts will vanish:

$$C_s(F) = C^{(ij)}\, E_i\, E_j + C_{(i+3,\, j+3)}\, B^i\, B^j. \tag{5.16}$$

**Theorem:**

$$F(\mathfrak{H}) = 0 \qquad \text{iff} \qquad C_s(F) = 0.$$

**Proof**: When $F$ and $\mathfrak{H}$ have the forms above in (5.7) and (5.10), respectively, one will gets:

$$F(\mathfrak{H}) = E(\mathbf{D}) - H(\mathbf{B}) = C^{ij}\, E_i\, E_j + C_{i+3,\, j+3}\, B^i\, B^j = C_s(F). \tag{5.17}$$

Commonly, when one assumes that the constitutive law is isotropic, in addition, one will have:

$$C^{ij} = \varepsilon\, \delta^{ij}, \qquad C_{i+3,\, j+3} = \frac{1}{\mu}\, \delta_{ij}, \tag{5.18}$$

in which $\varepsilon$ is the dielectric constant ([1]) of the medium and $\mu$ is its magnetic permeability.

This will make:

$$F(\mathfrak{H}) = \varepsilon\, \mathbf{E}^2 - \frac{1}{\mu}\, \mathbf{B}^2, \tag{5.19}$$

which is an expression that typically appears in the Lagrangian for an electromagnetic field, although the right-hand side is quite restrictive in scope when compared to the left-hand side.

One of the properties of the fields of electromagnetic waves is that they must lie on both the Klein quadric and the quadric that is associated with the constitutive law. Thus:

$$V(F) = 0, \qquad C_s(F) = 0. \tag{5.20}$$

which can also be written:

$$F \wedge F = 0, \qquad F(\mathfrak{H}) = 0. \tag{5.21}$$

Hence, such fields must be associated with lines in $\mathbb{R}P^3$ that belong to a certain quadratic line complex that is defined by the constitutive law.

**6. Summary.** We shall summarize the basic results of this article in the form of a table that translates electromagnetic concepts into their corresponding line-geometric ones, and which follows on the next page.

---

([1]) Of course, it will only be constant for an electrically-homogeneous medium.

Table 1. Electromagnetic concepts vs. line-geometric ones

| **Electromagnetism** | **Line geometry** |
|---|---|
| $\mathfrak{H} = \partial_t \wedge \mathbf{D} + \#_s^{-1} H$ | Pair of lines in $\mathbb{R}\mathrm{P}^3$: |
| $\partial_t \wedge \mathbf{D}$ | Line to infinity (point at infinity) |
| $\#_s^{-1} H$ | Line at infinity |
| $H(\mathbf{D}) = \mathbf{H} \cdot \mathbf{D} = 0$ | $\mathfrak{H} \wedge \mathfrak{H} = 0$: Klein quadric in $\mathrm{P}\Lambda_2$ |
| $F = dt \wedge E - \#_s\mathbf{B}$ | Linear complex |
| $dt \wedge E$ | Line at infinity |
| $\#_s\mathbf{B}$ | Line to infinity |
| rank $F = 2$ | Special complex |
| rank $F = 4$ | Null correlation: $\mathbb{R}\mathrm{P}^3 \to \mathbb{R}\mathrm{P}^{3*}$ |
| $E(\mathbf{B}) = \mathbf{E} \cdot \mathbf{B} = 0$ | $F \wedge F = 0$: dual Klein quadric in $\mathrm{P}\Lambda_2$ |
| Linear, non-dispersive, electromagnetic constitutive law | Polarity on $\mathrm{P}\Lambda^2$,<br>Correlation $[C] : \mathrm{P}\Lambda^2 \to \mathrm{P}\Lambda_2$ |
| $0 = F(\mathfrak{H}) = E(\mathbf{D}) - H(\mathbf{B})$ | Quadric of that polarity |
| Necessary conditions for electromagnetic waves:<br>$F \wedge F = 0$, $F(\mathfrak{H}) = 0$ | Quadratic line complex |

## References (*)

1. F. Klein, "Vergleichende Betrachtungen über neuere geometrische Forschungen," Habilitationsschrift, Erlanger, 1872, *Gesammelte mathematische Abhandlungen* XXVII; Eng. trans. by M. W. Haskell, "A comparative review of recent researches in geometry," Bull. New York Math. Soc. **2** (1892-1893), 215-249.
2. J. Plücker, *Neue Geometrie des Raumes*, B. G. Teubner, Leipzig, 1868.
3. F. Klein, *Nicht-Euklidische Geometrie*, Chelsea, NY, 1927.
4. D. M. Y. Sommerville, *Analytical Geometry of Three Dimensions*, Cambridge University Press, Cambridge, 1847.
5. J. G. Semple and G. T. Kneebone, *Algebraic Projective Geometry,* Clarendon Press, Oxford, 1952.
6. H. Busemann and P. J. Kelley, *Projective Geometry and Projective Metrics*, Academic Press, N. Y., 1953; reprinted by Dover, Mineola, NY, 2006.

(*) References marked with an asterisk are available in English translation as free downloads at the author's website: neo-classical-physics.info.

7. A. L. Onishchik and R. Sulanke, *Projective and Cayley-Klein Geometries*, Springer, Berlin, 2006.
8. G. Birkhoff, *Lattice Theory*, Amer. Math. Soc., Providence, R. I., 1940.
9*. F. Klein, "Über Liniengeometrie und metrische Geometrie," Math. Ann. **5** (1872); Ges. math. Abh. VIII, pp. 106-126.
10. C. M. Jessup, *A Treatise on the Line Complex*, Cambridge University Press, Cambridge, 1903; reprinted, with corrections, by A.M.S. Chelsea Publishing, Providence, R. I., 1969.
11. P. Gschwind, *Die lineare Komplex – eine überimaginäre Zahl*, Philosophisch-Anthroposophischer Verlag am Goetheanum, Dornach, 1991.
12*. F. Klein:

– "Notiz, betreffend den Zusammenhang der Liniengeometrie mit der Mechanik starren Körper," Math. Ann. 4 (1871); Ges. math. Abh., XIV, pp. 226-238.

– "Zur Schraubentheorie von Sir Robert Ball," Zeit. Math. u. Physik, **47** (1902); republished with an appendix in the Math. Ann., **62** (1906); Ges. math. Abh. XXIX, pp. 525-554.

13. E. Study, *Geometrie der Dynamen*, B. G. Teubner, Leipzig, 1903.
14. Sir R. Ball, *A treatise on the Theory of Screws*, Cambridge University Press, Cambridge, 1900.
15. O. Bottema and B. Roth, *Theoretical Kinematics*, North Holland, Amsterdam, 1979; reprinted by Dover, Mineola, N. Y., 1990.
16. J. M. McCarthy, *Introduction to Theoretical Kinematics*, M.I.T. Press, Cambridge, MA, 1990.
17. J. M. Selig, *Geometric Fundamentals of Robotics*, 2nd ed., Springer, Berlin, 2005.
18. D. H. Delphenich, "Projective geometry and special relativity," Ann. Phys. (Leipzig) **15** (2006), 216-246.
19*. F. Klein, "Über die geometrischen Grundlagen der Lorentzgruppe," Jber. der Deutschen Math.-Ver. **19** (1910); Ges math. Abh. XXX, pp. 533-552.
20. P. Gschwind:

– *Methodische Grundlagen zu einer projektiven Quantenphysik*, Philosophisch-Anthroposophischer Verlag am Goetheanum, Dornach, 1989.

– *Projektive Mikrophysik*, Philosophisch-Anthroposophischer Verlag am Goetheanum, Dornach, 2004.

21. D. H. Delphenich, *Pre-metric Electromagnetism*, free E-book, available at neo-classical-physics.info.
22 F. W. Hehl and Y. N. Obukhov, *Foundations of Classical Electrodynamics*, Birkhäuser, Boston, 2003.